\documentclass[preprint]{ptephy_v2}
\usepackage[colorlinks=true, linkcolor=blue, citecolor=blue, urlcolor=blue]{hyperref}
\usepackage{amsmath,amsthm,amssymb,amsfonts,bm}
\usepackage{graphicx}
\usepackage{color}
\usepackage{physics}

\newcommand{\inv}[1]{\frac{1}{#1}}

\begin{document}

\title{Analyzing the Higgs-confinement transition \\ with non-local operators on the lattice}
\author{Yusuke~Shimada}
\author{Arata~Yamamoto}
\affil{Department of Physics, The University of Tokyo, Tokyo 113-0033, Japan}

\begin{abstract}
We study non-local operators for analyzing the Higgs-confinement phase transition in lattice gauge theory.
Since the nature of the Higgs-confinement phase transition is topological, its order parameter is the expectation value of non-local operators, such as loop and surface operators.
There exist several candidates for the non-local operators.
Adopting the charge-2 Abelian Higgs model, we test numerical simulation of conventional ones, the Polyakov loop and the 't Hooft loop, and an unconventional one, the Aharonov-Bohm phase defined by the Wilson loop wrapping around a vortex line.
\end{abstract}

\subjectindex{B01}
\preprintnumber{2501.10662}

\maketitle

\section{Introduction}

In gauge theories coupled with fundamental matter, the Higgs phase and the confinement phase cannot be distinguished by any local order parameter because they have the same symmetry in the Ginzburg-Landau sense \cite{Fradkin:1978dv,Banks:1979fi,Osterwalder:1977pc}.
When the theory has topological order, the two phases are distinguishable by non-local order parameters and separated by a first-order phase transition.
Otherwise, the two phases are continuously connected.
A prominent example of the Higgs-confinement continuity is quantum chromodynamics (QCD) at nonzero density.
Quarks form color singlet hadrons in the vacuum while quarks form diquark Cooper pairs with color-flavor locking in high density limit.
The hadronic phase and the color-flavor locked phase have the same symmetry breaking pattern, and thus are expected to be smoothly connected without phase boundary.
This is referred to as the quark-hadron continuity \cite{Schafer:1998ef}.
The quark-hadron continuity might be relevant for phenomenology of dense QCD matter, such as neutron star physics \cite{Baym:2017whm}.

Non-local operators, such as loop and surface operators, play a key role in the Higgs-confinement phase transition.
Well-known examples of the non-local operators are the Polyakov and 't Hooft loops.
These loops are (dis)order parameters of electric and magnetic confinement, respectively.
In general, there is more than one order parameter for one phase transition.
Among several candidates, a topological vortex \cite{Alford:2018mqj,Chatterjee:2018nxe,Cherman:2018jir,Hirono:2018fjr} and its Aharonov-Bohm phase \cite{Cherman:2020hbe,Hayashi:2023sas} were intensely discussed in the context of the quark-hadron continuity.
Also, it was recently suggested that a correlation function on a vortex can distinguish the Higgs and confinement phases even if the transition is a smooth crossover \cite{Hayata:2024nrl}.
Numerical simulation of lattice gauge theory is a powerful way to compute these non-local operators nonperturbatively.
Some of them were already established and others were not.
We need to deepen our understanding of their lattice implementation.

In this paper, we study the lattice implementation of a few non-local operators for analyzing the Higgs-confinement phase transition.
We aim to showcase numerical analysis and to discuss practical difficulty etc.
For this purpose, we should choose a simple gauge-Higgs model as a testing ground.
We adopt the charge-$q$ Abelian Higgs model introduced by Fradkin and Shenker long ago~\cite{Fradkin:1978dv} because its phase structure is well understood, thanks to many earlier works~\cite{Creutz:1979he,Bowler:1981cj,Callaway:1981rt,Ranft:1982hf,Matsuyama:2019lei,Bhanot:1981ug,Azcoiti:1986wp,Matsuyama:2020tvt,Bonati:2024sok}.
Once we understand the implementation in this simple model, we can proceed to the next stage in forthcoming works; e.g., the SU(3) gauge-Higgs model, which is the Ginzburg-Landau theory of dense QCD.

The rest of the paper is organized as follows. 
We begin with a brief introduction of the Abelian Higgs model and the dual lattice in Sec.~\ref{secmodel}.
We discuss three non-local operators: the Polyakov loop in Sec.~\ref{secP}, the 't Hooft loop in Sec.~\ref{secT}, and the Aharonov-Bohm phase in Sec.~\ref{secAB}.
All the equations are written in the lattice unit throughout the paper.

\section{Abelian Higgs model}
\label{secmodel}

We introduce the lattice formulation of the charge-$q$ Abelian Higgs model in $3+1$ dimensions.
We employ the compact formulation with the unit-radius Higgs field.
In the compact formulation, there are only two kinds of field variables: the Higgs field angle $\theta (x)$ and the Abelian gauge field $A_\mu (x)$.
The action takes the form
\begin{equation}
\begin{split}
    S =& -\beta \sum_{\mu<\nu} \sum_x \cos F_{\mu\nu}(x) \\ 
    & - \kappa \sum_\mu \sum_x \cos \{ qA_\mu(x)-\theta(x)+\theta(x+\hat{\mu} ) \} \,,
    \label{eq:action}
\end{split}
\end{equation}
where $\hat{\mu}$ is the unit lattice vector in $\mu$ direction and $F_{\mu\nu}(x)$ is the discretized field strength
\begin{equation}
    F_{\mu\nu}(x) = A_\nu (x+\hat{\mu}) - A_\nu (x) - A_\mu (x+\hat{\nu}) + A_\mu (x) \, .
\end{equation}
This action is invariant under the local U(1) gauge transformation 
\begin{align}
    A_\mu '(x) &= A_\mu(x) + \lambda(x) - \lambda(x+\hat{\mu}) \,, \\
    \theta '(x) &= \theta(x) + q\lambda(x) \,
\end{align}
by an arbitrary function $\lambda(x)$ and the global $Z_q$ transformation
\begin{equation}
    A_\mu '(x) = A_\mu(x) + \frac{2\pi}{q} .
\end{equation}
The global $Z_q$ symmetry exists only for multiple charge $q\ge 2$.
We set $q=2$ in our numerical simulation.

The model has two parameters: the gauge coupling constant $\beta$ and the hopping parameter $\kappa$.
These parameters control the phase of matter.
The small $\beta$ regime is the confinement phase.
The fields are strongly coupled and electrically charged particles are confined. 
The large $\beta$ and large $\kappa$ regime is the Higgs phase.
Since the hopping term is dominant, $\theta(x)$ is frozen and Higgs condensation is induced.
The large $\beta$ and small $\kappa$ regime is the Coulomb phase.
The vacuum is the weakly-coupled perturbative one.
For $q\ge 2$, these three regimes are separated by the first-order phase transitions.
In particular, the Higgs and confinement regimes are distinguishable by the global $Z_q$ symmetry.
This is in stark contrast to the $q=1$ model, where the global $Z_q$ symmetry is absent and the confinement and Higgs regimes are smoothly connected without phase transition.
The phase diagram of the charge-$q$ Abelian Higgs model was investigated in several previous lattice calculations.
The calculation was first done with $Z_N$ approximation~\cite{Creutz:1979he}, and then with exact U(1) for the specific heat~\cite{Bowler:1981cj}, hysteresis~\cite{Callaway:1981rt}, and magnetic monopole flow~\cite{Ranft:1982hf}. 
The location of phase boundaries was recently obtained with large lattice sizes~\cite{Matsuyama:2019lei}.
The $(2+1)$-dimensional model was also studied~\cite{Bhanot:1981ug,Bonati:2024sok}.

We also introduce the dual lattice for later convenience.
The above model is defined on a four-dimensional hypercubic lattice with the volume $V=N_xN_yN_zN_\tau$.
The dual lattice is also a four-dimensional hypercubic lattice with the volume $V$.
The dual sites are defined by the geometric centers of hypercubes of the original lattice.
A $n$-dimensional manifold on the original lattice has one-to-one correspondence to a $(4-n)$-dimensional manifold on the dual lattice.
We use a star symbol $*$ for describing dual manifolds.
For example, an one-dimensional link $\mathcal{L}$ penetrates a three-dimensional dual cube $\mathcal{V}^*$, a two-dimensional face $\mathcal{S}$ intersects a two-dimensional dual face $\mathcal{S}^*$, and a three-dimensional cube $\mathcal{V}$ is penetrated by an one-dimensional dual link $\mathcal{L}^*$.
Magnetic objects, such as magnetic monopoles and vortices, are not elemental variables in the path integral but defined on the dual manifolds.
The operators to create the magnetic objects are defined so as to act on the corresponding manifolds on the original lattice.
The explicit operator constructions are demonstrated in Secs.~\ref{secT} and \ref{secAB}. 
Note that the Abelian Higgs model can be rewritten and simulated by using dual variables \cite{DelgadoMercado:2012tte,DelgadoMercado:2013ybm,Hayata:2019rem}, but we do not adopt such dual formulation in this work.

\section{Polyakov loop}
\label{secP}

The Polyakov loop is the most established order parameter of the confinement-deconfinement phase transition~\cite{Polyakov:1975rs}. 
Since the Polyakov loop is zero in the confinement phase and nonzero in the deconfinement phase, it is sometimes called a disorder parameter.
The Polyakov loop was calculated in many studies of lattice gauge theory, mainly in SU($N$) gauge theory. 
As for the charge-$q$ Abelian Higgs model, there is only one calculation and brief description~\cite{Matsuyama:2019lei}.
Here we provide more detailed analysis.

\begin{figure}[ht]
    \centering
    \begin{tabular}{ccc}
         \begin{minipage}{.32\linewidth}
            \centering
            \includegraphics[width = \linewidth]{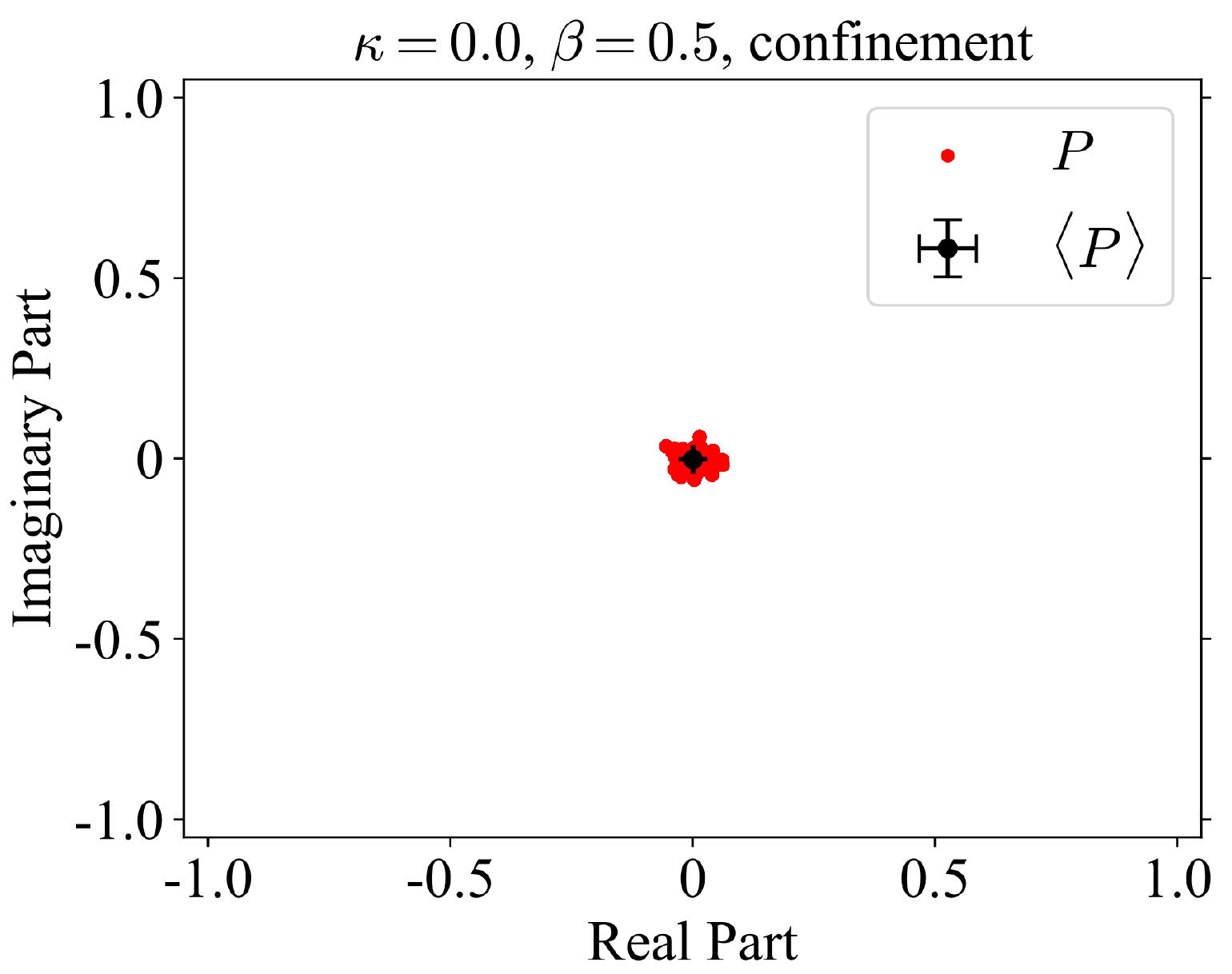}
         \end{minipage}&
         \begin{minipage}{.32\linewidth}
            \centering
            \includegraphics[width = \linewidth]{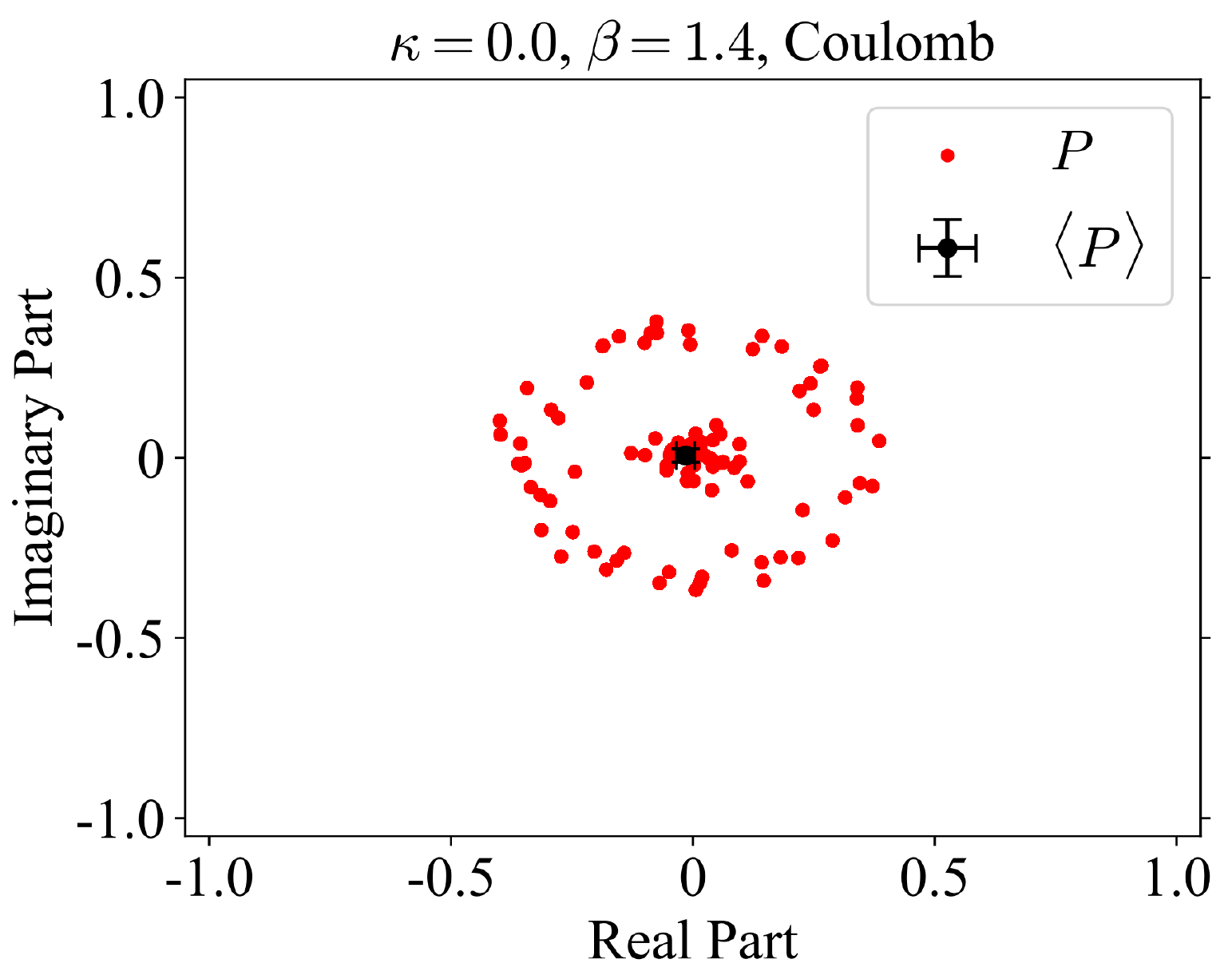}
         \end{minipage}&
         \begin{minipage}{.32\linewidth}
            \centering
            \includegraphics[width = \linewidth]{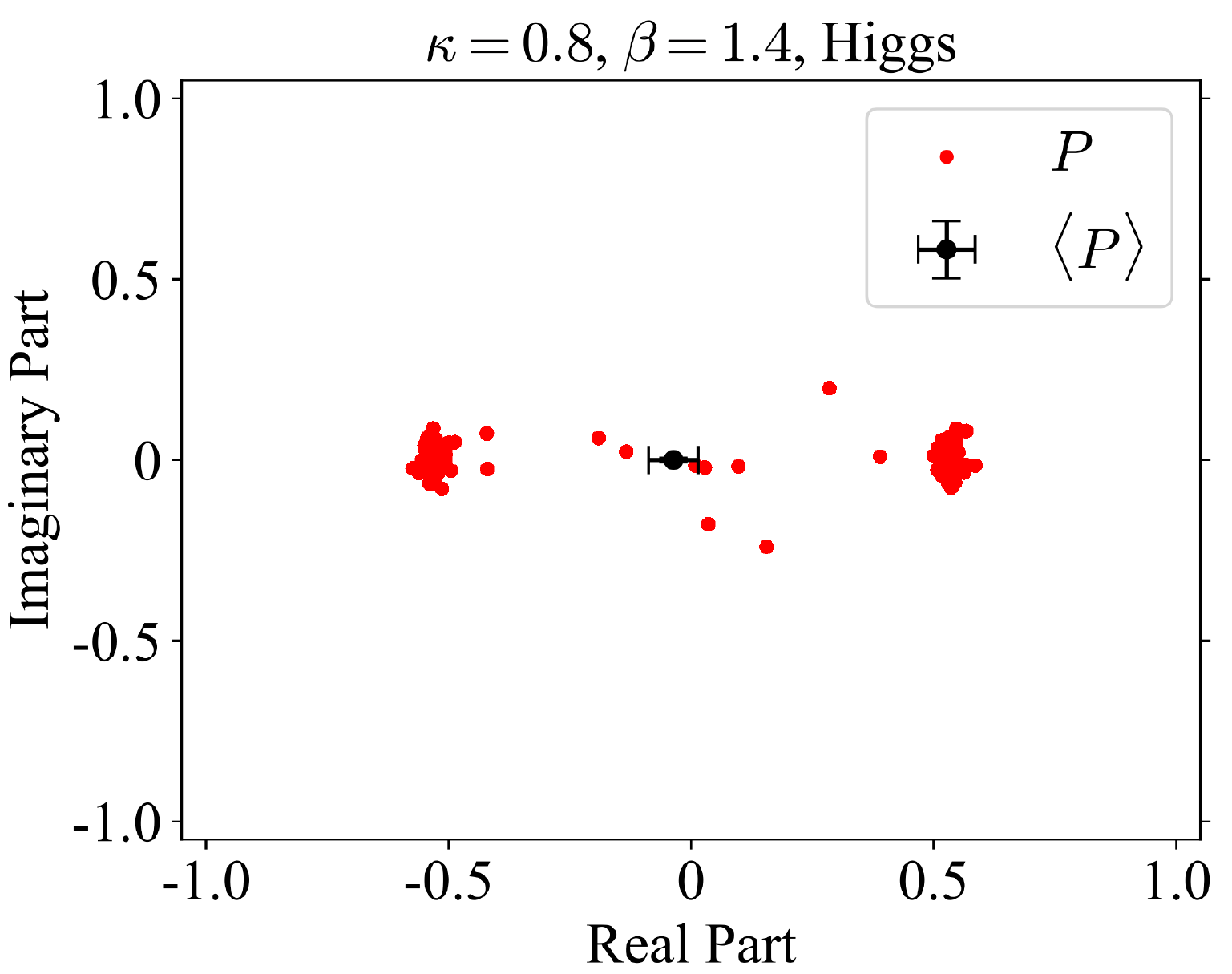}
         \end{minipage}
    \end{tabular}
    \caption{\label{fig_Poldistribution}
    Scattering plots of the averaged Polyakov loop $P$ and its expectation value $\expval{P}$. 
    The three panels have different parameters: $\kappa = 0.0, \beta = 0.5$ (left), $\kappa = 0.0, \beta = 1.4$ (center), and $\kappa = 0.8, \beta = 1.4$ (right).
    The lattice volume is $V=10^4$. Error bars are statistical.
    }
\end{figure}

We define the averaged Polyakov loop
\begin{equation}
    P = \inv{N_xN_yN_z} \sum_{\bm{r}} p(\bm{r})
    \label{eq:P}
\end{equation}
from the Polyakov loop at the spatial coordinate $\bm{r}$
\begin{equation}
    p(\bm{r}) = \prod_\tau \exp \{i A_4(\bm{r},\tau)\} \, .
\end{equation}
The expectation value
\begin{equation}
    \langle P \rangle = \frac{\int DA D\theta \ P e^{-S}}{\int DA D\theta \ e^{-S}}
\end{equation}
and $\langle p(\bm{r}) \rangle$ are order parameters of spontaneous $Z_q$ symmetry breaking.
In numerical simulation, however, they are zero regardless of the phase because symmetry is not spontaneously broken in finite volume.
The difference of the phase is seen in probability distribution.
Figure~\ref{fig_Poldistribution} is scattering plots of the averaged Polyakov loop for each configuration.
The number of gauge configuration is $N_{\mathrm{conf}} = 100$.
The left panel is the case of $\kappa = 0$ and $\beta = 0.5$, which corresponds to the confinement phase of pure U(1) gauge theory.
The distribution is very close to zero.
For large $\beta$, the system goes to the Coulomb phase.
The distribution is away from zero and spreads into a circle, as shown in the middle panel.
The circular distribution reflects the global U(1) symmetry in the pure gauge theory.
The right panel corresponds to the Higgs phase. 
For $\kappa>0$, the hopping term in Eq.~\eqref{eq:P} breaks the global symmetry from U(1) to $Z_2$.
When $\kappa$ is very large, the hopping term restricts link variables to $\pm 1$ and the model is reduced to $Z_2$ gauge theory. 
The distribution is concentrated in two minima on the real axis.
The change of the distribution from U(1) to $Z_2$ characterizes the Coulomb-Higgs phase transition.

\begin{figure}[ht]
    \centering
    \begin{tabular}{ccc}
         \begin{minipage}{.45\linewidth}
            \centering
            \includegraphics[width = \linewidth]{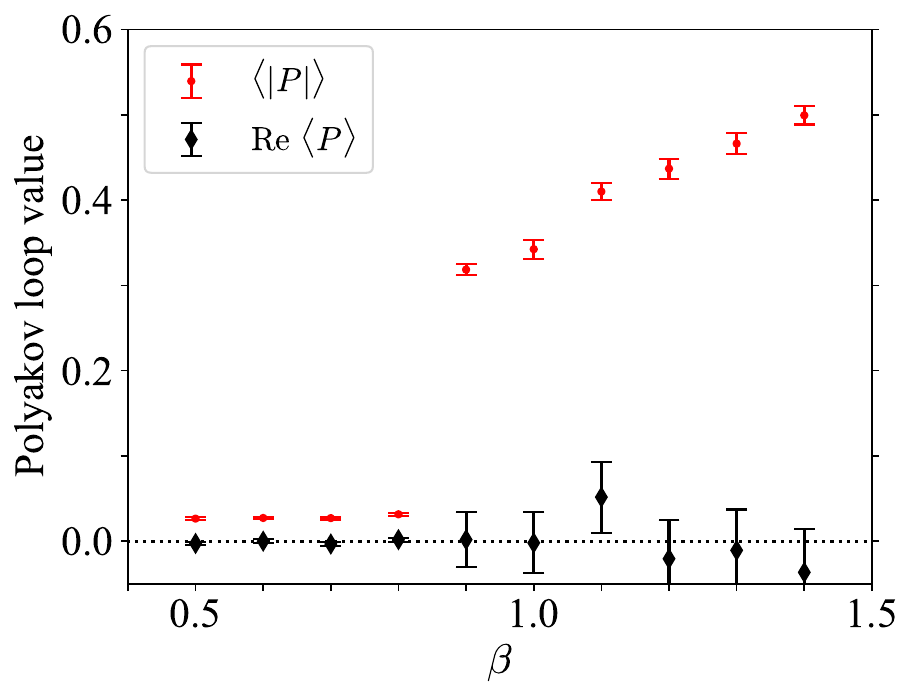}
         \end{minipage}&
         \begin{minipage}{.45\linewidth}
            \centering
            \includegraphics[width = \linewidth]{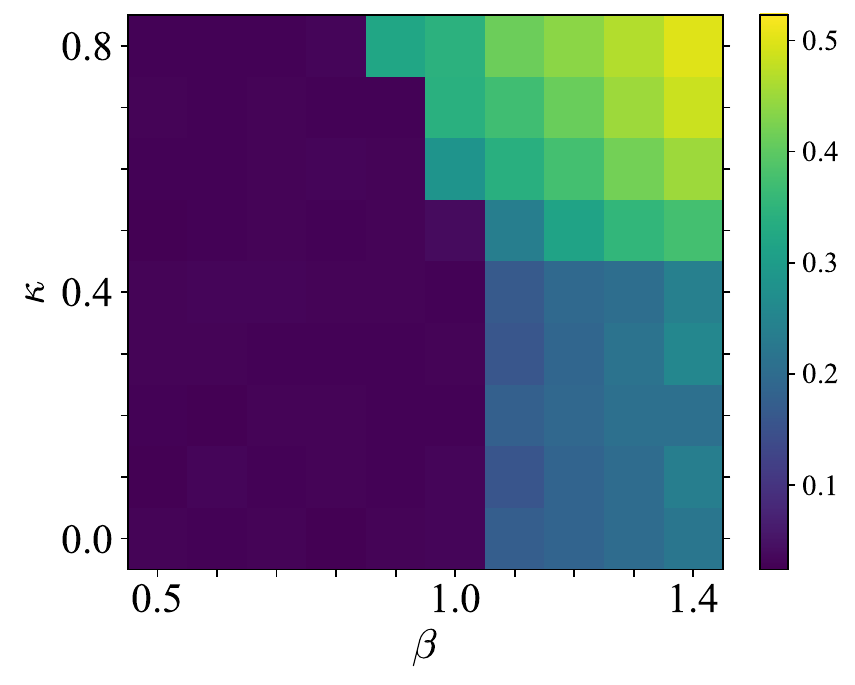}
         \end{minipage}
    \end{tabular}
    \caption{
    \label{fig_Polvalmap}
    Left: the expectation values of the averaged Polyakov loop $\langle P \rangle$ and its absolute value $\langle |P| \rangle$. 
    The hopping parameter is $\kappa=0.8$ and the lattice volume is $V=10^4$. Error bars are statistical.
    Right: a color plot of $\langle |P| \rangle$ in $\beta$-$\kappa$ plane with $\beta = 0.5, 0.6, \cdots, 1.4$ and $\kappa = 0.0, 0.1, \cdots, 0.8$. The dark region on the left side is the confinement phase and the light region on the right side is the deconfinement phase. The upper side of the deconfinement phase is the Higgs phase and the lower side is the Coulomb phase.
    }
\end{figure}

For determining transition points and drawing a phase diagram, we use $\expval{\abs{P}}$ instead of $\expval{P}$.
(The non-averaged one $\expval{p(\bm{r})}$ cannot be used for this purpose because of $|p(\bm{r})|=1$ in U(1) gauge theory. 
This is the reason why we take the spatial average in Eq.~\eqref{eq:P}.)
The results are shown in Fig.~\ref{fig_Polvalmap}. 
The left panel shows the development of the Polyakov loop against $\beta$ at fixed $\kappa$. 
While $\expval{P}$ is always zero, $\expval{\abs{P}}$ is nonzero and changes from small to large values.
There is a sharp jump between $\beta=0.8$ and $\beta=0.9$.
This is the confinement-deconfinement phase transition and its critical coupling constant is $0.8<\beta_c<0.9$ at $\kappa = 0.8$.
The right panel is the phase diagram in $\beta$-$\kappa$ plane, where the values of $\expval{\abs{P}}$ are highlighted in different colors. 
The phase boundaries of the three phases are clearly seen.
At $\kappa = 0$, the system is pure U(1) gauge theory and the critical coupling constant is $\beta_c \simeq 1.0$, which is consistent with pure gauge simulation~\cite{Creutz:1979zg,DeGrand:1980eq,Lautrup:1980xr}. 
The critical coupling constant $\beta_c$ gradually decreases as $\kappa$ increases.
The deconfinement phase is further decomposed into the Higgs and Coulomb regimes.
They can be distinguished by the values of the Polyakov loop; the Higgs regime has larger values and the Coulomb regime has smaller values.
This can be understood from the different distribution in Fig.~\ref{fig_Poldistribution}. 
The Higgs-Coulomb phase transition occurs at $\kappa_c = 0.4$-$0.5$. 
All the results are consistent with previous studies~\cite{Bowler:1981cj, Callaway:1981rt, Ranft:1982hf, Matsuyama:2019lei}.

\section{'t Hooft loop}
\label{secT}

The 't Hooft loop is dual to the Polyakov loop \cite{tHooft:1977nqb}.
(In this paper, we consider the 't Hooft loop encircling imaginary time direction.)
The 't Hooft loop is the worldline of a magnetic monopole while the Polyakov loop is the worldline of an electric charge.
Since the 't Hooft loop is nonzero in the confinement phase and zero in the Higgs phase, it is an order parameter of the Higgs-confinement phase transition.
 
\begin{figure}[ht]
\begin{minipage}[h]{0.48\linewidth}
\begin{center}
\includegraphics[width=0.8\textwidth]{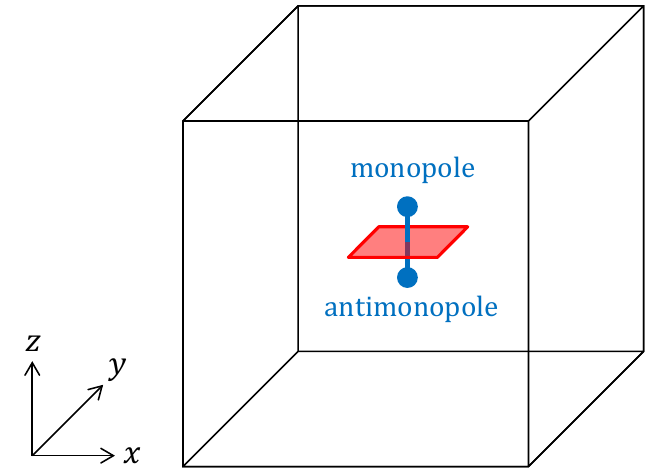}
\end{center}
\end{minipage}
\begin{minipage}[h]{0.48\linewidth}
\begin{center}
 \includegraphics[width=1\textwidth]{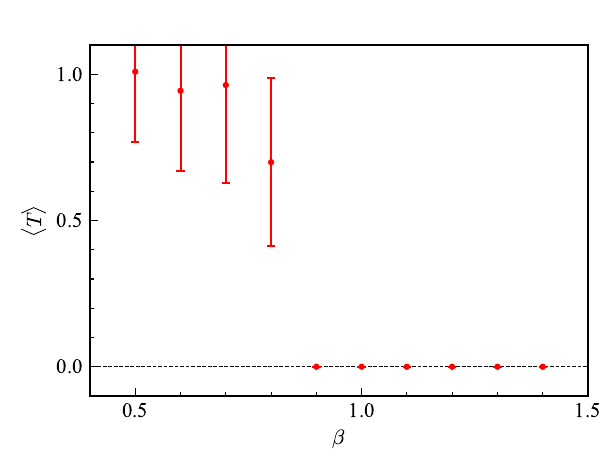}
\end{center}
\end{minipage}
\caption{\label{figV}
 Left: a schematic figure of a monopole-antimonopole pair.
 The red plaquette is the two-dimensional surface $\mathcal{S}$.
 Boundary conditions are periodic.
Right: the expectation value of 't Hooft loop $\langle T \rangle$.
The hopping parameter is $\kappa=0.8$ and the lattice volume is $V=10^4$.
Error bars are statistical.
}
\end{figure}

The lattice implementation of the 't Hooft loop has been known since long ago \cite{Mack:1978kr,Ukawa:1979yv,Srednicki:1980gb} and applied to SU($N$) gauge theory~\cite{Billoire:1981ye,DeGrand:1981yx,Kovacs:2000sy,Hart:2000hq,Hoelbling:2000su,DelDebbio:2000cb,deForcrand:2000fi,deForcrand:2001nd,deForcrand:2005pb,deForcrand:2005zg}.
The implementation generates the worldline of a magnetic monopole in $2+1$ dimensions.
Here we apply it to the Abelian gauge theory in $3+1$ dimensions.
In $(3+1)$-dimensional case, the implementation generates the worldsheet of a magnetic string, not the worldline of a single monopole, because of extra spatial direction, i.e., $z$ direction.
Let us consider the worldsheet on a two-dimensional dual surface $\mathcal{S}^*$ in $z$-$\tau$ plane.
The surface $\mathcal{S}^*$ has unit length in $z$ direction and wraps around $\tau$ direction.
The corresponding magnetic string has unit length and its endpoints are identified as a monopole and an antimonopole, as shown in the left panel of Fig.~\ref{figV}.
We define the surface operator $T$ to create the worldsheet.
The operation of $T$ multiplies a nontrivial $Z_q$ element to all the plaquette variables intersecting $\mathcal{S}^*$.
The expectation value of $T$ is given by 
\begin{equation}
    \langle T \rangle = \frac{\int DA D\theta \ T e^{-S}}{\int DA D\theta \ e^{-S}} = \frac{\int DA D\theta \ e^{-S'}}{\int DA D\theta \ e^{-S}} = \frac{\int DA D\theta \ e^{-S} e^{-(S'-S)}}{\int DA D\theta \ e^{-S}} = \langle e^{-(S'-S)} \rangle
    \label{eqT1}
\end{equation}
with the modified action
\begin{equation}
\begin{split}
    S' =& -\beta \sum_{x,\mu<\nu \notin \mathcal{S}} \cos F_{\mu\nu}(x) - \beta \sum_{x,\mu<\nu \in \mathcal{S}} \cos \left\{ F_{\mu\nu}(x) + \frac{2\pi}{q} \right\} \\ 
    & - \kappa \sum_{x, \mu} \cos \{ qA_\mu(x)-\theta(x)+\theta(x+\hat{\mu} ) \} .
    \label{eqSp}
\end{split}
\end{equation}
The surface $\mathcal{S}$ is defined as a set of the $x$-$y$ plaquettes intersecting $\mathcal{S}^*$.
The expectation value is positive real.
It is interpreted as $\exp(-F_{m})$, where $F_{m}$ is the free energy of a unit-length magnetic string, i.e., a monopole-antimonopole pair.

As shown in the right panel of Fig.~\ref{figV}, the simulation results clearly exhibit a jump of the first-order phase transition in spite of finite volume.
The free energy of the monopole-antimonopole pair (two monopoles for $q=2$) is finite in the confinement phase and nearly infinite in the Higgs phase.
The transition point $\beta_c = 0.8$-$0.9$ is consistent with that estimated from the Polyakov loop in the previous section.
This indicates that the above formulation works well for analyzing the Higgs-confinement phase transition.
Strictly speaking, however, the expectation value is not an order parameter of any symmetry.
In $3+1$ dimensions, it is not the free energy of a single monopole but a monopole-antimonopole pair.
The monopole-antimonopole pair has the same symmetry as the vacuum and its free energy will be large but finite even if a monopole is confined.
The expectation value seems nearly zero but is not exactly zero in Fig.~\ref{figV}.

In practice, we computed Eq.~\eqref{eqT1} by the Monte Carlo sampling with sequential reweighting \cite{deForcrand:2000fi}.
The computational cost of the reweighting method rapidly increases as the area of $\mathcal{S}$ increases.
The area of $\mathcal{S}$ is proportional to $N_\tau$ and the length of the magnetic string in $z$ direction.
Above we considered the magnetic string with unit length because the computational cost is small.
The magnetic string wrapping around $z$ direction, i.e., without endpoints, can be considered and its free energy can be computed.
This will show similar tendency although the cost is larger than ours.

\section{Aharonov–Bohm phase}
\label{secAB}

The difference between the Higgs and confinement regimes also emerges in nontrivial correlation between non-local operators.
It was recently conjectured that the two regimes are distinguished by the Aharonov–Bohm phase around a superfluid vortex \cite{Cherman:2020hbe}.
This conjecture was analytically examined in the strong coupling and deep Higgs limits \cite{Hayashi:2023sas}.
Lattice simulation will be necessary for full quantum examination.
Here we formulate the lattice simulation of the Aharonov–Bohm phase and test it in the Abelian Higgs model.
Note that the vortex in the Abelian Higgs model is not a global vortex of superfluid but a gauged vortex of superconductor and that the Aharonov–Bohm phase is common in both regimes of this model.
Although physics is different, the model can be used for testing formulation.
The formulation is straightforwardly applicable to lattice gauge theories with a superfluid vortex, such as the non-Abelian Higgs model~\cite{Yamamoto:2018vgg}.

\begin{figure}[ht]
\begin{minipage}[h]{0.48\linewidth}
\begin{center}
 \includegraphics[width=0.8\textwidth]{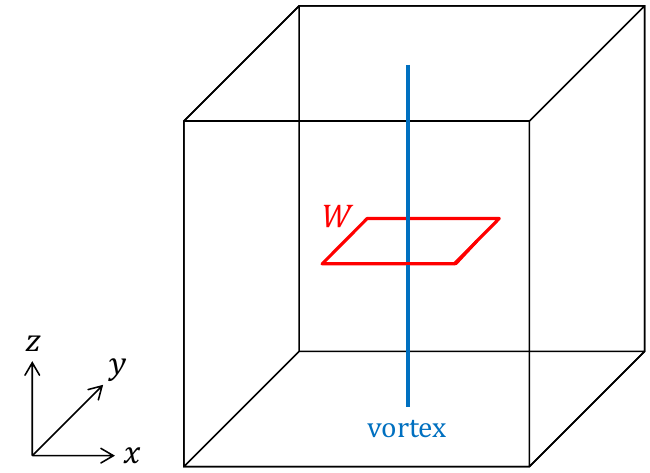}
\end{center}
\end{minipage}
\begin{minipage}[h]{0.48\linewidth}
\begin{center}
 \includegraphics[width=1\textwidth]{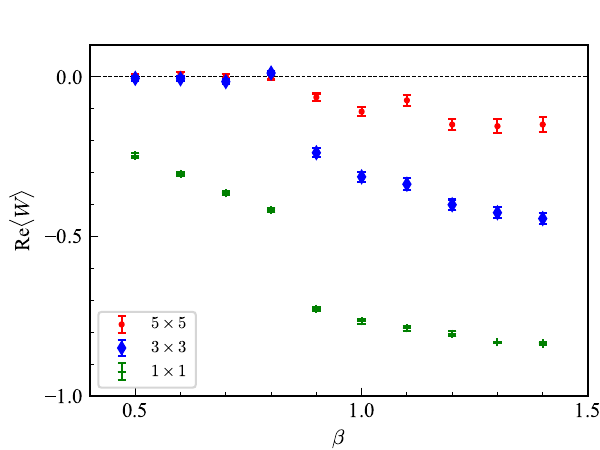}
\end{center}
\end{minipage}
 \caption{\label{figW}
 Left: a schematic figure of the spatial Wilson loop around a vortex.
 Boundary conditions are periodic.
 Right: the real part of the spatial Wilson loop $\langle W \rangle$ around a vortex.
 The hopping parameter is $\kappa=0.8$ and the lattice volume is $V=10^4$.
 Error bars are statistical.}
\end{figure}

A vortex of the Higgs field is defined by $2\pi$ circulation of $\theta(x)$.
In the compact formulation, $2\pi$ circulation of $\theta(x)$ is equivalent to $2\pi/q$ shift of $F_{\mu\nu}(x)$ through redefinition of the fields.
Thus the action results in Eq.~\eqref{eqSp}.
Now the surface $\mathcal{S}^*$ is the worldsheet of the vortex.
The setup is schematically depicted in the left panel of Fig.~\ref{figW}.
The worldsheet wraps around both $z$ and $\tau$ directions.
We define the spatial Wilson loop
\begin{equation}
    W = \prod_{x,\mu\in \mathcal{C}} \exp\{ iA_\mu(x) \}
\end{equation}
encircling the vortex, as shown in the figure.
The path $\mathcal{C}$ is a square on the $x$-$y$ plane and its center is located at the vortex position.
We computed the expectation value
\begin{equation}
\label{eqW}
    \langle W \rangle = \frac{\int DA D\theta \ e^{-S'} W}{\int DA D\theta \ e^{-S'}} .
\end{equation}
Since the spatial Wilson loop is the trajectory of a charged particle, it picks up the Aharonov–Bohm phase when it encircles the vortex.
We took the average of the translationally invariant $N_zN_\tau$ loops in Eq.~\eqref{eqW}.

The right panel of Fig.~\ref{figW} shows the real parts of the $1 \times 1$, $3 \times 3$, and $5 \times 5$ Wilson loops.
The imaginary parts are always zero in the case of $q=2$.
The results are clear in the Higgs phase.
Negative expectation values correspond to nonzero Aharonov–Bohm phase $\pi$ under coherent Higgs condensation.
A discontinuous change is seen at $\beta_c = 0.8$-$0.9$.
The loop size dependence of the Wilson loop changes from the perimeter law in the deconfinement regime to the area law in the confinement regime.
The exponential damping by the area law makes the signal to noise ratio much worse for larger loop.
In $\beta\le 0.8$, the expectation values of $3 \times 3$ and $5 \times 5$ loops are comparable with statistical error and the Aharonov–Bohm phase is not calculable.
Large statistics will be necessary to read off the large loop limit of the Aharonov–Bohm phase in the confinement phase.

\begin{figure}[ht]
\begin{minipage}[h]{0.48\linewidth}
\begin{center}
 \includegraphics[width=1\textwidth]{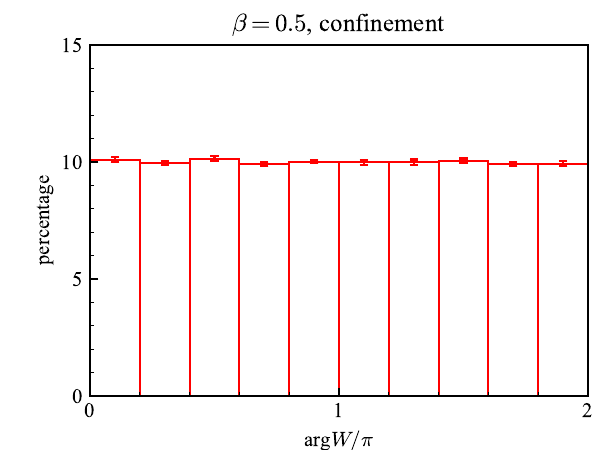}
\end{center}
\end{minipage}
\begin{minipage}[h]{0.48\linewidth}
\begin{center}
 \includegraphics[width=1\textwidth]{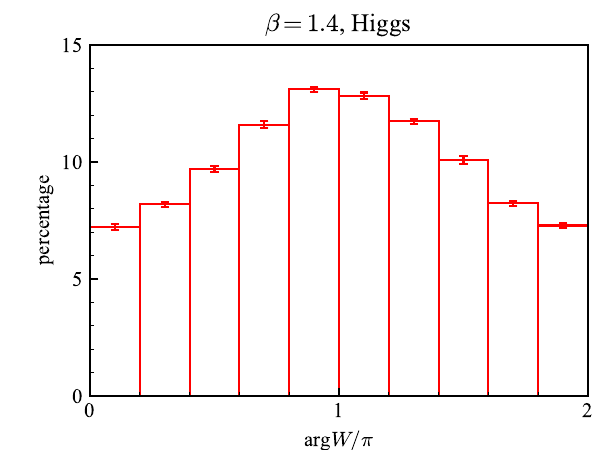}
\end{center}
\end{minipage}
 \caption{\label{figH}
 Probability distribution of the Aharonov–Bohm phase ${\rm arg} W$ at $\beta=0.5$ (left) and $\beta=1.4$ (right).
 The size of the spatial Wilson loop is $5\times 5$, the hopping parameter is $\kappa=0.8$, and the lattice volume is $V=10^4$.}
\end{figure}

For looking more closely at the Aharonov–Bohm phase, let us quantify the angle of the Wilson loop.
The expectation value $\langle {\rm arg} W \rangle$ is however ill-defined in quantum theory because it depends on the domain of argument.
For example, when ${\rm arg} W$ fluctuates around $\pi$, $\langle {\rm arg} W \rangle \simeq \pi$ if the argument is defined in $(0,2\pi]$, but $\langle {\rm arg} W \rangle \simeq 0$ if the argument is defined in  $(-\pi,\pi]$.
Instead, we here introduce probability distribution of ${\rm arg} W$.
Because of $|W|=1$ in U(1) gauge theory, probability distribution of ${\rm arg} W$ uniquely determines the expectation value.
We generated $N_{\rm conf}=100$ gauge configurations, computed $N_zN_\tau=10^2$ translationally invariant Wilson loops, and made one histogram, i.e., how many data exist in each domain of ${\rm arg} W$, from $100 \times 10^2$ data.
We repeated these steps for 10 independent sets of gauge configurations to evaluate statistical average and error.
Thus, totally, we made a histogram from the $100 \times 10^2 \times 10 = 10^5$ data.
As seen from the histograms in Fig.~\ref{figH}, the probability distribution changes between the Higgs and confinement regimes.
In the confinement regime, the phase is random because of strong fluctuation of gauge fields.
In the Higgs regime, the probability distribution has a maximum at $\pi$, but it is not very peaky.
This means that quantum fluctuation is still strong and semi-classical approximation is invalid.
The probability distribution will go to the delta function in classical limit, such as the deep Higgs limit $\kappa\to\infty$.

\ack
The authors thank Yui Hayashi for fruitful discussion.
This work was supported by JSPS KAKENHI Grant No.~19K03841 and No.~24KJ0658, and JST SPRING Grant No.~JPMJSP2108. 
The numerical simulations were partly carried out on SX-ACE in Osaka University.

\bibliographystyle{apsrev4-2}
\bibliography{paper}

\end{document}